\documentclass[aps,floats,prd,showpacs,fancyhdr,a4paper,superscriptaddress,nofootinbib]{revtex4}
\usepackage{latexsym}
\usepackage{amsmath}
\usepackage{amssymb}
\usepackage{amsfonts}
\usepackage{graphicx}

\usepackage{dcolumn}
\usepackage{fancyhdr}

\makeatletter
\def\onpage#1#2#3{\ifnum\c@page=#1 #2\else#3\fi}
\def\pagesbefore#1#2#3{\ifnum\c@page<#1 #2\else#3\fi}
\def\pagesafter#1#2#3{\ifnum\c@page>#1 #2\else#3\fi}
\makeatother

\begin{document}


\title{Full Causal Bulk Viscous Cosmological Models with Variable $G$ and $\Lambda$}
\author{J. A. Belinch\'on}
\email{abelcal@ciccp.es}
\affiliation{\small {Grupo Inter-Universitario de An\'{a}lisis Dimensional.\\ Dept. F\'{\i}sica ETS Arquitectura UPM Av. Juan de Herrera 4 Madrid 28040 Espa\~na}}

\author{T. Harko } 
\email{tcharko@hkusua.hku.hk}
\affiliation{Department of Physics, The University of Hong Kong,
Pokfulam Road, Hong Kong, P. R. China.}
\author{M. K. Mak}
\email{mkmak@vtc.edu.hk}
\affiliation{Department of Physics, The Hong Kong University of Science and Technology,Clear Water Bay, Hong Kong, P. R. China.}

\date{\today}

\begin{abstract}
We study the evolution of a flat Friedmann-Robertson Walker Universe, filled
with a bulk viscous cosmological fluid, in the presence of variable
gravitational and cosmological constants. The dimensional analysis of the
model suggest a proportionality between the bulk viscous pressure of the
dissipative fluid and the energy density. With the use of this assumption and
with the choice of the standard equations of state for the bulk viscosity
coefficient, temperature and relaxation time, the general solution of the
field equations can be obtained, with all physical parameters having a
power-law time dependence. The symmetry analysis of this model, performed by
using Lie group techniques, confirms the unicity of the solution for this
functional form of the bulk viscous pressure.
\end{abstract}
\pacs{98.80.Hw, 98.80.Bp, 04.20.Jb, Variable Constants, Bulk Viscous models, Cosmology}
\maketitle

\section{Introduction}

In many cosmological and astrophysical situations an idealised fluid model of
the matter is inappropriate. Such possible situations are the relativistic
transport of photons, mixtures of cosmic elementary particles, the evolution
of cosmic strings due to their interaction with each other and with the
surrounding matter, the classical description of the (quantum) particle
production phase, interaction between matter and radiation, quark and gluon
plasma viscosity, different components of dark matter etc. \cite{ChJa96}. From
a physical point of view the inclusion of dissipative terms in the energy
momentum tensor of the cosmological fluid seems to be the best-motivated
generalization of the matter term of the gravitational field equations.

The theories of dissipation in Eckart-Landau formulation \cite{Ec40},
\cite{LaLi87}, who made the first attempt at creating a relativistic theory of
viscosity, are now known to be pathological in several respects. Regardless of
the choice of equation of state, all equilibrium states in these theories are
unstable. In addition, as shown by Israel \cite{Is76}, signals may be
propagated through the fluid at velocities exceeding the speed of light. These
problems arise due to the first order nature of the theory since it considers
only first-order deviations from the equilibrium, leading to parabolic
differential equations, hence to infinite speeds of propagation for heat flow
and viscosity, in contradiction with the principle of causality. While such
paradoxes appear particularly glaring in relativistic theory, infinite
propagation speeds already constitutes a difficulty at the classical level,
since one does not expect thermal disturbances to be carried faster than some
(suitably defined) mean molecular speed. Conventional theory is thus
applicable only to phenomena which are ``quasi-stationary'' i.e. slowly
varying on space and time scales characterized by mean free path and mean
collision time \cite{Is76}. This is inadequate for many phenomena in
high-energy astrophysics and relativistic cosmology involving steep gradients
or rapid variations. These deficiencies can be traced to the fact that the
conventional theories (both classical and relativistic) make overly
restrictive hypothesis concerning the relation between the fluxes and
densities of entropy, energy and particle number.

A relativistic second-order theory was found by Israel \cite{Is76} and
developed by Israel and Stewart \cite{IsSt76} into what is called `transient'
or \ `extended' irreversible thermodynamics. In this model deviations from
equilibrium (bulk stress, heat flow and shear stress) are treated as
independent dynamical variables leading to a total of $14$ dynamical fluid
variables to be determined. The solutions of the full causal theory are well
behaved for all times. Hence the advantages of the causal theories are the
followings \cite{AnPaRo98}: 1) for stable fluid configurations the dissipative
signals propagate causally 2) unlike Eckart-type's theories, there is no
generic short-wavelength secular instability in causal theories and 3) even
for rotating fluids, the perturbations have a well-posed initial value
problem. Therefore, the best currently available theory for analyzing
dissipative processes in the Universe is the full Israel-Stewart causal thermodynamics.

Due to the complicated nonlinear character of the evolution equations, very
few exact cosmological solutions of the gravitational field equations are
known in the framework of the full causal theory. For a homogeneous Universe
filled with a full causal viscous fluid source obeying the relation $\xi
\sim\rho^{\frac{1}{2}}$, exact general solutions of the field equations have
been obtained in \cite{ChJa97}, \cite{MaHa98}, \cite{MaTr97}-\cite{PrHeIb01}.
In this case the evolution of the bulk viscous cosmological model can be
reduced to a Painleve-Ince type differential equation. It has also been
proposed that causal bulk viscous thermodynamics can model on a
phenomenological level matter creation in the early Universe \cite{MaHa98},
\cite{MaHa99a}.

Recent observations of type Ia supernovae with redshift up to about
$z\lesssim1$ provided evidence that we may live in a low mass-density
Universe, with the contribution of the non-relativistic matter (baryonic plus
dark) to the total energy density of the Universe of the order of
$\ \Omega_{m}\sim0.3$ \cite{Ri98}-\cite{Pe98}. The value of $\Omega_{m}$ is
significantly less than unity \cite{OsSt95} and consequently either the
Universe is open or there is some additional energy density $\rho$ sufficient
to reach the value $\Omega_{total}=1$, predicted by inflationary theory.
Observations also show that the deceleration parameter of the Universe $q$ is
in the range $-1\leq q<0$, and the present-day Universe undergoes an
accelerated expansionary evolution.

Several physical models have been proposed to give a consistent physical
interpretation to these observational facts. One candidate, and maybe the most
convincing one for the missing energy is vacuum energy density or cosmological
constant $\Lambda$ \cite{We89}.

Since the pioneering work of Dirac \cite{Di38}, who proposed, motivated by the
occurrence of large numbers in Universe, a theory with a time variable
gravitational coupling constant $G$, cosmological models with variable $G$ and
nonvanishing and variable cosmological term have been intensively investigated
in the physical literature \cite{Ra88}-\cite{Wa93}. In the isotropic
cosmological model of Chen and Wu \cite{ChWu90} it is supposed, in the spirit
of quantum cosmology, that the effective cosmological constant $\Lambda$
varies as $a^{-2}$ (with $a$ the scale function). In the cosmological model of
Lima and Maia \cite{LiMa94} the cosmological constant $\Lambda=\Lambda\left(
H\right)  =3\beta H^{2}+3\left(  1-\beta\right)  H^{3}/H_{I}$ is a complicated
function of the Hubble parameter $H$, a constant $\beta$ and an arbitrary time
scale $H_{I}^{-1}$, leading to a cosmic history beginning from an instability
of the de Sitter space-time. The cosmological implications of a time
dependence of the cosmological of the form $\Lambda\sim t^{-2}$ have been
considered by Berman \cite{Be91}. Waga \cite{Wa93} investigated flat
cosmological models with the cosmological term varying as $\Lambda
=\alpha/a^{2}+\beta H^{2}+\gamma$, with $\alpha$, $\beta$ and $\gamma$
constants. In this model exact expressions for observable quantities can be
obtained. Nucleosynthesis in decaying-vacuum cosmological models based on the
Chen-Wu ansatz \cite{ChWu90} has been investigated by Abdel-Rahman
\cite{Ra92}. The consistency with the observed helium abundance and baryon
asymmetry allows a maximum vacuum energy close to the radiation energy today.
Anisotropic Bianchi type I cosmological models with variable $G$ and $\Lambda$
have been analyzed by Beesham \cite{Be93a} and it was shown that in this case
there are no classical inflationary solutions of pure exponential form.
Cosmological models with the gravitational and cosmological constants
generalized as coupling scalars and with $G\sim a^{n}$ have been discussed by
Sistero \cite{Si91}. Generalized field equations with time dependent $G$ and
$\Lambda$ have been proposed in \cite{La85} and \cite{LaPr86} in an attempt to
reconcile the large number hypothesis with Einstein's theory of gravitation.
Limits on the variability of $G$ using binary-pulsar data have been obtained
by Damour, Gibbons and Taylor \cite{DaGiTa88}. A detailed analysis of
Friedmann-Robertson-Walker (FRW) Universes in a wide range of scalar-tensor
theories of gravity has been performed by Barrow and Parsons \cite{BaPa97}.

It is the purpose of the present paper to consider the evolution of a causal
bulk viscous fluid filled flat FRW type Universe, by assuming the standard
equations of state for the bulk viscosity coefficient, temperature and
relaxation time and in the presence of variable gravitational and cosmological
constants. In order to obtain some very general properties of this
cosmological model with variable constants we shall adopt a method based on
the studies of the symmetries of the field equations. As a first step we shall
study the field equations from dimensional point of view. The dimensional
method provides general relations between physical quantities and allow to
make some definite assumptions on the behavior of thermodynamical quantities.
In particular we find that, under the assumption of the conservation of the
total energy of the Universe, the bulk viscous pressure of the cosmological
fluid must be proportional to the energy density of the matter component. With
the use of this assumption the gravitational field equations can be integrated
exactly, leading to a general solution in which all thermodynamical quantities
have a power-law time dependence.

As we have been able to find a solution through dimensional analysis, it is
possible that there are other symmetries of the model, since dimensional
analysis is a reminiscent of scaling symmetries, which obviously are not the
most general form of symmetries. Hence we shall study the model through the
method of Lie group symmetries, showing \ that under the assumed hypotheses of
the proportionality of bulk viscous pressure to the energy density, there are
no other solutions of the field equations.

The present paper is organized as follows. The field equations are written
down in Section II. The dimensional analysis of the model is performed in
Section III. The general solution of the gravitational field equations for the
bulk viscous pressure proportional to the energy density is obtained in
Section IV. The Lie group symmetry study of the model is considered in Section
V. In Section VI we discuss and conclude our results.

\section{Field equations, thermodynamics and consequences}

In the presence of a time variable gravitational and cosmological constants
$G$ and $\Lambda$ the Einstein gravitational field equations are:
\begin{equation}
R_{ik}-\frac{1}{2}g_{ik}R=\frac{8\pi G(t)}{c^{4}}T_{ik}+\Lambda(t)g_{ik}.
\label{ECU1}%
\end{equation}

Applying the covariance divergence to the second member of equation
(\ref{ECU1}) one obtain the generalized ``conservation equation''
\begin{equation}
T_{i;j}^{j}=\left(  -\frac{G_{,j}}{G}\right)  T_{i}^{j}-\frac{c^{4}%
\Lambda_{,i}}{8\pi G}.
\end{equation}

We assume that the geometry of the space-time is of FRW type, with a line
element
\begin{equation}
ds^{2}=c^{2}dt^{2}-a^{2}(t)\left(  dx^{2}+dy^{2}+dz^{2}\right)  . \label{line}%
\end{equation}

The Hubble parameter associated with the line element (\ref{line}) is defined
as $H=\frac{\dot{a}}{a}$.

The energy-momentum tensor of the bulk viscous cosmological fluid filling the
very early Universe is given by
\begin{equation}
T_{i}^{k}=\left(  \rho c^{2}+p+\Pi\right)  u_{i}u^{k}-\left(  p+\Pi\right)
\delta_{i}^{k},\label{1}%
\end{equation}
where $\rho$ is the mass density, $p$ the thermodynamic pressure, $\Pi$ the
bulk viscous pressure and $u_{i}$ the four velocity satisfying the condition
$u_{i}u^{i}=1$. The particle and entropy fluxes are defined according to
$N^{i}=nu^{i}$ and $S^{i}=eN^{i}-\left(  \frac{\tau\Pi^{2}}{2\xi T}\right)
u^{i}$, with $n$ is the number density, $e$ the specific entropy, $T\geq0$ the
temperature, $\xi$ the bulk viscosity coefficient and $\tau\geq0$ the
relaxation coefficient for transient bulk viscous effect (i.e. the relaxation time).

The evolution of the cosmological fluid is subject to the dynamical laws of
particle number conservation $N_{\text{ };i}^{i}=0$ and Gibb's equation
$Tde=d\left(  \frac{\rho c^{2}}{n}\right)  +pd\left(  \frac{1}{n}\right)  $.
In the following we shall also suppose that the energy-momentum tensor of the
cosmological fluid is conserved, that is $T_{i;k}^{k}=0$.

For the evolution of the bulk viscous pressure we adopt the causal evolution
equation \cite{Ma95}, obtained in the simplest way (linear in $\Pi)$ to
satisfy the $H$-theorem (i.e. for the entropy production to be non-negative,
$S_{;i}^{i}=\frac{\Pi^{2}}{\xi T}\geq0$ \cite{IsSt76}$).$

Therefore the gravitational field equations describing the cosmological
evolution of a causal bulk viscous fluid in the presence of variable
gravitational and cosmological constants are%

\begin{align}
3H^{2} &  =8\pi G(t)\rho+\Lambda(t)c^{2},\label{field1}\\
2\dot{H}+3H^{2} &  =-\frac{8\pi G(t)}{c^{2}}\left(  p+\Pi\right)
+\Lambda(t)c^{2},\label{field2}\\
\dot{\rho}+3\left(  \rho+\frac{p}{c^{2}}+\frac{\Pi}{c^{2}}\right)  H &
=0,\label{field3}\\
8\pi\rho\dot{G}(t)+\dot{\Lambda}(t)c^{2} &  =0,\\
\tau\dot{\Pi}+\Pi &  =-3\xi H-\frac{\epsilon}{2}\tau\Pi\left(  3H+\frac
{\dot{\tau}}{\tau}-\frac{\dot{\xi}}{\xi}-\frac{\dot{T}}{T}\right)
.\label{field4}%
\end{align}

In eq. (\ref{field4}), $\epsilon=0$ gives \ the truncated theory (the
truncated theory implies a drastic condition on the temperature), while
$\epsilon=1$ gives the full theory. The non-causal theory has $\tau=0$. Dot
denotes differentiation with respect to time.

In order to close the system of equations (\ref{field1}-\ref{field4}) we have
to give the equation of state for $p$ and specify $T$, $\xi$ and $\tau$. As
usual, we assume the following phenomenological laws \cite{Ma95}:
\begin{equation}
p=(\gamma-1)\rho c^{2},\text{ \ }\xi=\alpha\rho^{s},\text{ \ \ }T=\beta
\rho^{r},\text{ \ \ }\tau=\xi\rho^{-1}=\alpha\rho^{s-1}, \label{csi1}%
\end{equation}
where $\alpha\geq0$, $\beta\geq0$ are dimensional constants and $1\leq
\gamma\leq2$, $s\geq0$ and $r\geq0$ are numerical constants. Eqs. (\ref{csi1})
are standard in cosmological models whereas the equation for $\tau$ is a
simple procedure to ensure that the speed of viscous pulses does not exceed
the speed of light. These are without sufficient thermodynamical motivation,
but in absence of better alternatives we shall follow the practice adopting
them in the hope that they will at least provide indication of the range of
possibilities. The temperature law is the simplest law guaranteeing positive
heat capacity.

With the use of equations of state (\ref{csi1}) the evolution equation of the
bulk viscous pressure becomes
\begin{equation}
\dot{\Pi}+\frac{1}{\alpha}\rho^{1-s}\Pi=-3\rho H-\frac{1}{2}\Pi\left[
3H-\left(  1+r\right)  \frac{\dot{\rho}}{\rho}\right]  . \label{eq11}%
\end{equation}

In the context of irreversible thermodynamics $p$, $\rho$, $T$ and the
particle number density $n$ are equilibrium magnitudes which are related by
equations of state of the form $\rho=\rho(T,n)$ and $p=p(T,n)$. From the
requirement that the entropy is a state function, we obtain the equation
$\left(  \frac{\partial\rho}{\partial n}\right)  _{T}=\frac{p/c^{2}+\rho}%
{n}-\frac{T}{n}\left[  \frac{\partial}{\partial T}\left(  p/c^{2}\right)
\right]  _{n}$. For the equations of state (\ref{csi1}) this relation imposes
the constraint $r=\frac{\gamma-1}{\gamma}$, so that $0\leq r\leq1/2$ for
$1\leq\gamma\leq2$, a range of values which is usually considered in the
physical literature \cite{ChJa97}.

The growth of the total comoving entropy $\Sigma$ over a proper time interval
$\left(  t_{0},t\right)  $ is given by \cite{Ma95}:
\begin{equation}
\Sigma(t)-\Sigma\left(  t_{0}\right)  =-\frac{3}{k_{B}}\int_{t_{0}}^{t}%
\frac{\Pi Ha^{3}}{T}dt, \label{M entropy}%
\end{equation}
where $k_{B}$ is the Boltzmann's constant.

The Israel-Stewart-Hiscock theory is derived under the assumption that the
thermodynamical state of the fluid is close to equilibrium, that is the
non-equilibrium bulk viscous pressure should be small when compared to the
local equilibrium pressure $\left|  \Pi\right|  <<p=\left(  \gamma-1\right)
\rho c^{2}$ \cite{Zi96}. If this condition is violated then one is effectively
assuming that the linear theory holds also in the nonlinear regime far from
equilibrium. For a fluid description of the matter, the condition ought to be satisfied.

An important observational quantity is the deceleration parameter $q=\frac
{d}{dt}\left(  \frac{1}{H}\right)  -1$. The sign of the deceleration parameter
indicates whether the model inflates or not. The positive sign of $q$
corresponds to ``standard'' decelerating models whereas the negative sign
indicates inflation.

By using the field equations we can express the gravitational constant in the
form
\begin{equation}
G(t)=\frac{3}{4\pi}\frac{H\dot{H}}{\dot{\rho}}.\label{g}%
\end{equation}

With the use of this expression for $G$ we obtain the cosmological constant in
the form
\begin{equation}
\Lambda(t)c^{2}=3H^{2}\left(  1-\frac{2\dot{H}\rho}{H\dot{\rho}}\right)  .
\label{l}%
\end{equation}

From the field equations we obtain for the derivative of the Hubble function
the alternative expression
\begin{equation}
\dot{H}=-\frac{4\pi G(t)}{c^{2}}\left(  \rho c^{2}+p+\Pi\right)  .
\label{hdot}%
\end{equation}

\section{Dimensional analysis of field equations}

In this section we show how writing the field equations in a dimensionless way
a particular solution of the field equations, corresponding to a specific
choice of the bulk viscous pressure, can be obtained.

The $\pi-monomia$ is the main object in dimensional analysis. It may be
defined as the product of quantities which are invariant under change of
fundamental units. $\pi-monomia$ are dimensionless quantities, their
dimensions are equal to unity. The dimensional analysis has structure of Lie
group \cite{Ca88}. The $\pi-monomia$ are invariant under the action of the
similarity group. On the other hand we must mention that the similarity group
is only a special class of the mother group of all symmetries that can be
obtained using the Lie method. For this reason when one uses dimensional
analysis only one of the possible solutions to the problem is obtained.

The equations (\ref{field1}-\ref{field4}) and the equation of state
(\ref{csi1}) can be expressed in a dimensionless way by the following
$\pi-monomia$:
\begin{align}
\pi_{1} &  =\frac{Gpt^{2}}{c^{2}}\text{ ,\ \ }\pi_{2}=\frac{G\left|
\Pi\right|  t^{2}}{c^{2}},\text{ \ }\pi_{3}=G\rho t^{2},\text{ \ }\pi
_{4}=\frac{\left|  \Pi\right|  }{p},\text{ \ }\pi_{5}=\frac{\xi}{\left|
\Pi\right|  t},\text{ \ }\pi_{6}=\frac{\tau}{t}=\tau H\text{\ },\label{pi1}\\
\text{ \ }\pi_{7} &  =\frac{\xi}{\alpha\rho^{s}},\text{ \ \ }\pi_{8}=\frac
{\xi}{\tau\rho},\text{ }\pi_{9}=\frac{T}{\beta\rho^{r}},\text{ \ }\pi
_{10}=\frac{\rho c^{2}}{p},\text{ \ }\pi_{11}=\Lambda c^{2}t^{2}.\label{pi4}%
\end{align}

The following relation can be obtained from the $\pi-monomia.$ From $\pi
_{1},\pi_{2},\pi_{3},\pi_{4}$ and $\pi_{10}$ we can see that
\begin{equation}
\rho\propto p\propto\left|  \Pi\right|  ,
\end{equation}
note that $\left[  \left|  \Pi\right|  \right]  =\left[  p\right]  =\left[
\rho c^{2}\right]  ,$ that is, they have the same dimensional equation and
therefore from this point of view they must have the same behavior in order of
magnitude. Now, from $\pi_{7}=\frac{\xi}{\alpha\rho^{s}}$ and $\pi_{8}%
=\frac{\xi}{\tau\rho}$ we obtain
\begin{equation}
\pi_{8}=\frac{\alpha\rho^{s-1}}{t},\label{edensity}%
\end{equation}
implying
\begin{equation}
\rho\sim\left(  \frac{t}{\alpha}\right)  ^{\frac{1}{s-1}},\text{ \ }s\neq1.
\end{equation}

From $\pi_{3}$ and $\pi_{8}$ we find
\begin{equation}
\frac{G\rho t^{2}}{c^{2}}=\frac{\alpha\rho^{s-1}}{t},\text{ \ \ \ } \label{cG}%
\end{equation}
or
\begin{equation}
\frac{\alpha^{\frac{1}{s-1}}c^{2}}{G}=t^{\frac{1-2s}{1-s}},s\neq1.
\end{equation}

If $s=1/2$ we obtain the relationship $\alpha^{2}=c^{2}/G.$ If $s\neq1/2$ the
``constant'' $G$ must vary.

Therefore the solutions that Dimensional Analysis suggests us are the
following ones:
\begin{align}
\rho &  \propto\left(  \alpha^{-1}t\right)  ^{\frac{1}{s-1}},\text{ }%
p\propto\left(  \alpha^{-1}t\right)  ^{\frac{1}{s-1}}=\left(  \gamma-1\right)
\rho c^{2},\text{ }\left|  \Pi\right|  \propto\left(  \alpha^{-1}t\right)
^{\frac{1}{s-1}}=\chi\rho c^{2}\\
\text{ }\Lambda &  \propto c^{-2}t^{-2},\text{ \ }G\propto\alpha^{\frac
{1}{s-1}}c^{2}t^{\frac{2s-1}{1-s}},\text{ \ }s\neq1,
\end{align}
with $\chi$ a numerical constant. From the equation of state of bulk viscosity
coefficient, temperature and relaxation time we find
\begin{equation}
\xi\propto\alpha\left(  \alpha^{-1}t\right)  ^{\frac{s}{s-1}},\text{
\ }T\propto\beta\left(  \alpha^{-1}t\right)  ^{\frac{r}{s-1}},\text{ \ \ }%
\tau\propto t,\text{ \ \ }s\neq1.
\end{equation}

\section{General solution of the gravitational field equations with bulk
viscous pressure proportional to the energy density}

As we have seen in the previous Section, dimensional analysis suggests us that
$\left|  \Pi\right|  \propto\rho c^{2}.$ Hence generally the bulk viscous
pressure can be represented in the form $\Pi=-\chi\rho c^{2}$, with $\chi
\geq0$ and $\chi\ll1.$ With this hypothesis the conservation equation becomes
\begin{equation}
\dot{\rho}+3\left(  \gamma-\chi\right)  \rho H=0,
\end{equation}
leading to the following relationship between the energy density $\rho$ and
the scale factor $a$:
\begin{equation}
\rho=\frac{\tilde{\rho}_{0}}{a^{3\left(  \gamma-\chi\right)  }}, \label{dens0}%
\end{equation}
where $\tilde{\rho}_{0}>0$ is an arbitrary constant of integration$.$ With the
use of the assumed functional form of $\Pi$ the evolution equation of the bulk
viscous pressure becomes
\begin{equation}
\dot{\rho}=-K_{0}\rho^{2-s},
\end{equation}
where we denoted $K_{0}=\frac{1}{\alpha}\left[  \frac{1-r}{2}+\frac{2-\chi
}{2\chi\left(  \gamma-\chi\right)  }\right]  .$ The solution of this equation
is%
\begin{equation}
\rho(t)=\frac{\rho_{0}}{t^{\frac{1}{1-s}}}, \label{dens1}%
\end{equation}
with $\rho_{0}=\left[  \left(  1-s\right)  K_{0}\right]  ^{1/(s-1)}$. From
Eqs. (\ref{dens0}) and (\ref{dens1}) we obtain the time dependence of the
scale factor%
\begin{equation}
a(t)=a_{0}t^{\frac{1}{3\left(  1-s\right)  \left(  \gamma-\chi\right)  }}.
\end{equation}

with $a_{0}=\left(  \frac{\tilde{\rho}_{0}}{\rho_{0}}\right)  ^{\frac
{1}{3\left(  \gamma-\chi\right)  }}.$

The Hubble parameter is
\begin{equation}
H(t)=\frac{H_{0}}{t},
\end{equation}
where $H_{0}=\frac{1}{3\left(  1-s\right)  \left(  \gamma-\chi\right)  }>0.$
The behavior of the gravitational and cosmological constant can be obtained
from Eqs. (\ref{g}) and (\ref{l}), respectively:
\begin{equation}
G(t)=\frac{G_{0}}{t^{\frac{1-2s}{1-s}}},\text{ \ \ \ \ \ }\Lambda\left(
t\right)  =\frac{\Lambda_{0}}{t^{2}},
\end{equation}
where $G_{0}=\frac{3}{4\pi}\frac{H_{0}^{2}\left(  1-s\right)  }{\rho_{0}}$ and
$\Lambda_{0}=3H_{0}^{2}\left(  2s-1\right)  /c^{2}$. The behavior of the bulk
viscosity coefficient, temperature and relaxation time is given by
\begin{equation}
\xi=\frac{\xi_{0}}{t^{\frac{s}{1-s}}},\text{ \ \ \ }T=\frac{T_{0}}{t^{\frac
{r}{1-s}}},\text{ \ \ \ }\tau=\tau_{0}t,
\end{equation}
where we denoted $\xi_{0}=\alpha\rho_{0}^{s}$, $T_{0}=\beta\rho_{0}^{r}$ and
$\tau_{0}=\xi_{0}/\rho_{0}$. The expression of $\tau$ is in agreement with the
one expected from a theoretical point of view, as argued in \cite{Ma95}, since
for a viscous expansion to be non-thermalizing we should have $\tau<t,$ for
otherwise the basic interaction rate for viscous effects could be sufficiently
rapid to restore the equilibrium as the fluid expands.

The deceleration parameter $q$ behaves as:
\begin{equation}
q=1/H_{0}-1=const.
\end{equation}

For values of the parameters $s,$ $\gamma$ and $\chi$ so that $0<H_{0}<1$ the
expansion of the Universe is non-inflationary, while for $H_{0}>1$ we obtain
an inflationary behavior.

The comoving entropy in this model is
\begin{equation}
\Sigma(t)-\Sigma\left(  t_{0}\right)  =\Sigma_{0}t^{\frac{1}{1-s}\left(
\frac{1}{\gamma-\chi}+r-1\right)  },
\end{equation}
where we denoted $\Sigma_{0}=\frac{3(1-s)\chi\rho_{0}H_{0}a_{0}^{3}}%
{k_{B}T_{0}\left(  \frac{1}{\gamma-\chi}+r-1\right)  }$.

The ratio of the bulk viscous and thermodynamic pressure is given by
\begin{equation}
\left|  \frac{\Pi}{p}\right|  =\left|  \frac{\chi}{\gamma-1}\right|  .
\end{equation}

Since $\chi$ is assumed to be a small number, the present model is
thermodynamically consistent for the matter equations of state of cosmological
interest, describing high-density cosmological fluids, when bulk viscous
dissipative effects are important.

We would like to point out that we have obtained the same results as in the
previous section, at least in order of magnitude. By direct integration of the
field equations we have been able of finding some of the numerical constants
describing the evolution of the Universe. This is a crucial issue at least for
the cosmological constant. With this complete solution it is observed that we
have the special case $s=1/2$ for which we obtain
\begin{equation}
G=const.\text{, \ \ }\Lambda=0,\text{ \ \ }\rho\propto t^{-2},\text{
\ \ }a\propto t^{\frac{2}{3\left(  \gamma-\chi\right)  }},...,\Sigma
(t)-\Sigma\left(  t_{0}\right)  \propto t^{2(\frac{1}{\gamma-\chi}+r-1)}.
\label{FRW results}%
\end{equation}
If for example we take $\gamma=4/3$ (radiation predominance) then
\begin{equation}
\Sigma(t)-\Sigma\left(  t_{0}\right)  \propto t^{2(\frac{3}{4-3\varkappa
}-\frac{3}{4})}. \label{INT entropy}%
\end{equation}
We therefore see that the $\chi-$parameter, i.e. the causal bulk effect,
perturb weakly the FRW perfect fluid solution. If $\chi=0$ we recover the
perfect fluid case, and then the comoving entropy is constant.

If $s\neq1/2$ then we obtain the following behavior for $G$ and $\Lambda$: if
$1/2<s<1$ then $G$ is a growing function on time while $\Lambda>0$, and
$\Lambda\propto t^{-2}$, if $s<1/2$ and $\Lambda<0$ or $s>1$ and $\Lambda<0,$
then $G$ is a decreasing function on time while $\Lambda\propto t^{-2}$.

These results agree with the ones obtained in a recent work \cite{RG} where we
studied through different routes, renormalization group (RG), dimensional
analysis and structural stability a full causal cosmological model. The main
result that we obtained in \cite{RG} was that only and only for $s=1/2$ we
obtain that the fixed point of the RG equation corresponds to a flat
non-viscous FRW universe, i.e. a perfect fluid case. The same result is
obtained under the standard stability analysis performed in this work, showing
in this way that only for the case $s=1/2$ the causal bulk viscous model
approximates in the long time limit the dynamics of a flat perfect fluid
filled FRW Universe.

\section{Lie symmetries of the model}

In this Section we are going to study the field equations through the symmetry
method. As we have indicated in the introduction dimensional analysis is just
a manifestation of scaling symmetry, but this type of symmetry is not the most
general form of symmetries. Therefore by studying the form of $G(t)$ for which
the equations admit symmetries, we hope to uncover new integrable models. But
we shall see that under the assumptions made we only obtain the previous
solutions obtained in the above sections.

We start again with the assumption $\Pi=-\chi\rho c^{2}$, with $\chi\geq0$.
The bulk viscosity evolution equation can then be rewritten in the alternative
form
\begin{equation}
\frac{1-r}{2}\frac{\dot{\rho}}{\rho}+\frac{1}{\alpha}\rho^{1-s}=3\left(
\frac{1}{\chi}-\frac{1}{2}\right)  H.
\end{equation}

Taking the derivative with respect to the time of this equation and with the
use of Eq. (\ref{hdot}) we obtain the following second order differential
equation describing the time variation of the density of the cosmological
fluid:
\begin{equation}
\ddot{\rho}=\frac{\dot{\rho}^{2}}{\rho}-D\rho^{1-s}\dot{\rho}+AG(t)\rho^{2},
\label{neweq}%
\end{equation}
where $D=\frac{2\left(  1-s\right)  }{\alpha\left(  1-r\right)  }>0$ and
$A=\frac{12\pi\left(  \gamma-\chi\right)  \left(  \chi-2\right)  }{\chi\left(
1-r\right)  }<0.$

Equation (\ref{neweq}) is of the general form.
\begin{equation}
\ddot{\rho}=f(t,\rho,\dot{\rho}),\text{ \ \ \ }%
\end{equation}
where $f(t,\rho,\dot{\rho})=\frac{\dot{\rho}^{2}}{\rho}-D\rho^{1-s}\dot{\rho
}+AG(t)\rho^{2}$.

We are going now to apply all the standard procedure of Lie group analysis to
this equation (see \cite{Ibragimov} for details and notation)

A vector field $X$
\begin{equation}
X=\xi(t,\rho)\partial_{t}+\eta(t,\rho)\partial_{\rho},
\end{equation}
is a symmetry of (\ref{neweq}) if%

\[
-\xi f_{t}-\eta f_{\rho}+\eta_{tt}+\left(  2\eta_{t\rho}-\xi_{tt}\right)
\dot{\rho}+\left(  \eta_{\rho\rho}-2\xi_{t\rho}\right)  \dot{\rho}^{2}%
-\xi_{\rho\rho}\dot{\rho}^{3}+
\]
\begin{equation}
+\left(  \eta_{\rho}-2\xi_{t}-3\dot{\rho}\xi_{\rho}\right)  f-\left[  \eta
_{t}+\left(  \eta_{\rho}-\xi_{t}\right)  \dot{\rho}-\dot{\rho}^{2}\xi_{\rho
}\right]  f_{\dot{\rho}}=0. \label{ber2}%
\end{equation}

By expanding and separating (\ref{ber2}) with respect to powers of $\dot{\rho
}$ we obtain the overdetermined system:
\begin{align}
\xi_{\rho\rho}+\rho^{-1}\xi_{\rho}  &  =0,\label{ber1}\\
\eta\rho^{-2}-\eta_{\rho}\rho^{-1}+\eta_{\rho\rho}-2\xi_{t\rho}+2D\xi_{\rho
}\rho^{1-s}  &  =0,\label{ber1_1}\\
2\eta_{t\rho}-\xi_{tt}+D\xi_{t}\rho^{1-s}-3A\xi_{\rho}G\rho^{2}-2\eta_{t}%
\rho^{-1}+D\left(  1-s\right)  \eta\rho^{-s}  &  =0,\label{ber3}\\
-A\xi\dot{G}\rho^{2}-2A\eta G\rho+\eta_{tt}+\left(  \eta_{\rho}-2\xi
_{t}\right)  AG\rho^{2}+D\eta_{t}\rho^{1-s}  &  =0, \label{ber4}%
\end{align}
Solving (\ref{ber1}-\ref{ber4}), we find that
\begin{equation}
\xi(\rho,t)=-m(1-s)t+b,\text{\ }\eta(\rho,t)=m\rho
\end{equation}
with the constraint
\begin{equation}
\frac{\dot{G}}{G}=\frac{(2s-1)m}{m(1-s)t-b},
\end{equation}
and $m,b$ are numerical constant.

Thus we have found all the possible forms of $G$ for which eq. (\ref{neweq})
admits symmetries. There are two cases with respect to the values of the
constant $m$, $m=0$ which correspond to $G=const.$ and $m\neq0$ which
correspond to
\begin{equation}
G=G_{1}\left[  m(1-s)t+b\right]  ^{\frac{2s-1}{1-s}}, \label{SYM G}%
\end{equation}
with $G_{1}$ a constant of integration. If $s=\frac{1}{2}$ then $G$ is a
constant. For this form of $G$ eq. (\ref{neweq}) admits a single symmetry
\begin{equation}
X=\left(  m(1-s)t-b\right)  \partial_{t}-\left(  m\rho\right)  \partial_{\rho
}. \label{SYM X}%
\end{equation}

The knowledge of one symmetry $X$ might suggest the form of a particular
solution as an invariant of the operator $X$, i.e. the solution of
\begin{equation}
\frac{dt}{\xi\left(  t,\rho\right)  }=\frac{d\rho}{\eta\left(  t,\rho\right)
} \label{ecu7}%
\end{equation}

This particular solution is known as an invariant solution (generalization of
similarity solution). In this case
\begin{equation}
\rho=\rho_{0}t^{-\frac{1}{1-s}}, \label{SYM density}%
\end{equation}
where for simplicity we have taken $b=0,$ and $\rho_{0}$ is a constant of integration.

We can apply a pedestrian method to try to obtain the same results. In this
way, taking into account dimensional considerations, from the eq.
(\ref{neweq}) we obtain the following relationships between density, time and
gravitational constant:
\begin{equation}
\alpha^{-1}(1-s)\rho^{1-s}t\backsimeq1,\text{ \ \ }AG\rho t^{2}\backsimeq1.
\label{sciama}%
\end{equation}

This last relationship is also known as the relation for inertia obtained by
Sciama. From these relationship we obtain
\begin{equation}
\rho\thickapprox A^{-1}G_{1}^{-1}\left[  m(1-s)t+b\right]  ^{\frac{1-2s}{1-s}%
}t^{-2}\thickapprox A^{-1}G_{1}^{-1}\left[  m(1-s)t\right]  ^{-\frac{1}{1-s}}.
\label{sciama density}%
\end{equation}
We can see that this result verifies the relation
\begin{equation}
\alpha^{-1}(1-s)\rho^{1-s}t\backsimeq1.\text{ \ \ }%
\end{equation}

Therefore the conditions $\Pi=-\chi\rho c^{2}$ together with the equation
(\ref{eq11}) are very restrictive. Hence we have shown that under these
assumptions there are no another possible solutions for the field equations.

\section{Conclusions}

In the present paper we have studied a causal bulk viscous cosmological model,
with bulk viscosity coefficient proportional to the energy density of the
cosmological fluid. This hypothesis is justified by the dimensional analysis
of the model. We have also assumed that the cosmological and gravitational
constants are functions of time. With the help these assumptions the general
solution of the gravitational equations can be obtained in an exact form,
leading to a power-law time dependence of the physical parameters on the
cosmological time. The unicity of the solution is also proved by the
investigation of the Lie group symmetries of the basic equation describing the
time variation of the mass density of the Universe.

In the considered model the evolution of the Universe starts from a singular
state, with the energy density, bulk viscosity coefficient and cosmological
constant tending to infinity. At the initial moment $t=0$ the relaxation time
is $\tau=0$. Generally the behavior of the gravitational constant shows a
strong dependence on the coefficient $s$ entering in the equation of state of
the bulk viscosity coefficient. For $s=1/2$, $G$ is a constant during the
cosmological evolution. From the singular initial state the Universe starts to
expand, with the scale factor $a$ an $s$-dependent function. Depending on the
value of the constants in the equations of state, the cosmological evolution
is non-inflationary ($q>0$ for $H_{0}<1$) or inflationary ($q<0$ for $H_{0}%
>1$). Due to the proportionality between bulk viscous pressure and energy
density, both inflationary and non-inflationary models are thermodynamically
consistent, with the ratio of $\Pi$ and $p$ also much smaller than $1$ in the
inflationary era. Generally, the cosmological constant is a decreasing
function of time. The present model is not defined for $s=1$, showing that in
the important limit of small densities an other approach is necessary.

Bulk viscosity is expected to play an important role in the early evolution of
the Universe, when also the dynamics of the gravitational and cosmological
constants could be different. Hence the present model, despite its simplicity,
can lead to a better understanding of the dynamics of our Universe in its
first moments of existence.

\end{document}